\documentclass[12pt]{article}

\usepackage[iso-8859-7]{inputenc}
\usepackage{epsfig}
\usepackage{graphicx}
\usepackage{epstopdf}
\usepackage{amsfonts}
\usepackage{amssymb}
\usepackage{amsthm}
\usepackage{amsmath}
\usepackage{multirow}

\usepackage{color}

\newcommand{\newc}{\newcommand}
\newc{\ra}{\rightarrow}
\newc{\lra}{\leftrightarrow}
\newc{\be}{\begin{equation}}
\newc{\ee}{\end{equation}}
\newc{\bs}{\begin{split}}
\newc{\es}{\end{split}}
\newc{\ba}{\begin{eqnarray}}
\newc{\ea}{\end{eqnarray}}
\newc{\ov}{\overline}
\newc{\pa}{\partial}
\newc{\D}{\Delta}

\newc{\nn}{\nonumber}
\begin{document}
	\begin{titlepage}
		
		\vspace*{0.7cm}

		\begin{center}
			{\large	\bf  $SL(2,7)$ Representations and their relevance to Neutrino Physics  }
			\\[12mm]
			G. Aliferis$^{a}$
					\footnote{E-mail: \texttt{aliferis@auth.gr}},
		G. K. Leontaris$^{b, c}$
					\footnote{E-mail: \texttt{leonta@uoi.gr}}
						and	N. D. Vlachos$^{a}$
									\footnote{E-mail: \texttt{vlachos@physics.auth.gr}
						}
			\\[-2mm]
			
		\end{center}
		\vspace*{0.50cm}
			\centerline{$^{a}$ \it
					 Department of Nuclear and Particle Physics}
				\centerline{\it  University of Thessaloniki	}
				\vspace*{0.15cm}
					\centerline{\it GR-54124 Thessaloniki, Greece	}
		\centerline{$^{b}$\it Physics Department,  University of Ioannina}
				\centerline{\it
					GR-45110 Ioannina, Greece	}
		\vspace*{0.2cm}
			\centerline{$^{c}$ \it Department of Physics, CERN}
							\centerline{\it
	CH-1211, Geneva 23, Switzerland 	}
				
		\vspace*{1.00cm}
		
		\begin{abstract}
			\noindent
    The investigation of the r\^ole of finite groups in flavor physics and particularly, in the interpretation of the neutrino data has
    been the subject of intensive research.  Motivated by this fact, in this  work we derive the three-dimensional unitary representations
    of the projective linear group $PSL_2(7)$. Based on the observation that the  generators of the group exhibit a latin square pattern,
    we use available computational packages on discrete algebra to determine the generic properties of the group elements.  We present
    analytical expressions and discuss several examples which reproduce the neutrino mixing angles  in accordance
    with the experimental data.

\end{abstract}
		
	\end{titlepage}
	
	\thispagestyle{empty}
	\vfill
	\newpage

	\section{Synopsis}
	
	   Abelian and non-abelian  discrete symmetries have been extensively used to impose constraints on the Yukawa lagrangian.
	   In model building they are often used to generate hierarchical structures in the fermion mass matrices, to eliminate
	    proton decay operators and suppress other  terms inducing unobserved processes.  The structure of the  neutrino mass matrix and the experimentally
	    measured mixing angles in particular, hint to the existence of an underlying non-Abelian flavor symmetry. In this
	    context, the neutrino mass matrix is assumed to be invariant under certain transformations of the discrete
	    group $D_f$ (for reviews see~\cite{Altarelli:2010gt,Ishimori:2010au,King:2013eh}).   Therefore, in unified theories of the
	    fundamental gauge interactions the symmetry of the effective model is expected to contain a non-abelian gauge group $G_{GUT}$
	    accompanied by a (non)-abelian discrete flavor symmetry $D_f$. In some string  scenarios  the  total effective symmetry
	    $G_{GUT}\times D_f$ is usually embedded in a higher unified group, such as $E_8$~\cite{Karozas:2014aha}.
	     For the most familiar  GUT symmetries such as $E_6, SO(10)$ and $SU(5)$,   the discrete group is  a subgroup
	     of $SU(3)$.  In the past, several of these cases have been considered, including those belonging to the chains
	      $S_n, A_n$ and possess triplet representations.  In the present work we  focus on some particular cases
	      of a  general class  of discrete symmetries.  These are  the special linear groups $SL_2(p)$~\cite{Tanaka67,Humphreys75}
	     and their corresponding projective ones, $PSL_2(p)$, with $p$ prime number.  From this class of groups
	     it is sufficient for our purposes to take  $p\le 7$ since these are the only cases where the resulting discrete
	     groups contain triplet representations 	   and at the same time are embedded in $SU(3)$.     Moreover,
     from the physics point of view, the rather interesting case of $PSL_2(7)$ which is a simple subgroup of $SU(3)$,
     is less explored  (see however~\cite{Luhn:2007yr,King:2009tj,deAdelhartToorop:2011re}), 	and it is our main focus in the present work.
     Since we demand invariance of the  neutrino mass matrix under certain	group actions,  a prerequisite for such an analysis
     is the explicit form of the group elements of  $PSL_2(7)$.
      However, while finding the representations of the elements and the multiplication tables of $A_3, A_4, A_5$ is relatively easy,
	  this becomes an onerous  process for $PSL_2(7)$.
	  	
	  	In a previous work~\cite{Floratos:2015jvu},  using the automorphisms of the discrete and 	finite Heisenberg group, the 3-dimensional
	  	representations of the $PSL_2(p)$ generators were constructed.
	  	It is our  purpose here to  systematically derive  the structure of all the elements of the three-dimensional
	  	representations  of $PSL_2(7)$. The explicit form of the three-dimensional unitary representations
	  	  would be a very useful tool in many physics applications including
	  	  their possible relevance to the structure of the neutrino mass matrix and lepton mixing angles. Focusing on  neutrino physics,
	      a usual approach is to  construct neutrino and charged lepton mass matrices  invariant under certain elements of the  assumed group.
	       Hence, a  systematic exploration of the specific properties  and in particular the characteristic mixing matrix of the neutrino sector
		   require such a methodology.

Finite groups were proposed long time ago as a possible symmetry to the peculiar neutrino hierarchy and
this  is our basic motivation for the present construction.  With many details found in reviews and other works,
here,  we only give  a brief description of our assumptions.   We consider a scenario where the charged lepton
and neutrino mass matrices are subject to constraints under the same or different subgroups of a covering parent discrete symmetry. This is
compatible, for example, with model building in string theory framework. Taking  for example the $SU(5)$ gauge theory,
in some F-theory framework, the  trilinear  Yukawa couplings for the various types of fields are realized at different
points of the internal manifold and they correspond to different symmetry enhancements of the $SU(5)$  singularity~\cite{Beasley:2008kw}.
For example, the $10\cdot 10\cdot 5$ coupling is realized at a `point' of the  compact manifold associated  with $E_6$ enhancement
and the $10\cdot\bar 5\cdot
\bar 5$ at an $SO(12)$ enhancement. Similarly, the corresponding  discrete symmetry associated with these points may differ,
although they could be subgroups  of the same covering discrete group.  We assume that this is the case for
$m_{\ell}$ and $m_{\nu}$ which are not realized at the same `point'. Then, we  consider that
 the neutrino mass matrix commutes with an element $A$ of a given discrete group
\be
[ m_{\nu},A]=0\,,\label{MAcom}
\ee
while a similar relation holds for the charged lepton mass matrix, as well. Then, the vanishing of the commutator (\ref{MAcom}) implies that both, $m_{\nu}$ and $A$ have
 a common system of eigenvectors, hence they define the diagonalising (mixing) matrix $V_{\nu}$. In other words, $  U_{\nu}^{\dagger} AU_{\nu}
 =A^{diag.}$ as well as $ U_{\nu}^{\dagger} m_{\nu}U_{\nu} = m_{\nu}^{diag.}$.
	
The layout of the article is as follows. In section  section 2 we summarize the basic steps of  the generators construction ~\cite{Floratos:2015jvu}
using the work of~\cite{BI}. In section 3, based on the observation that the $PSL_2(7)$ 3-d representations exhibit a specific structure,  we calculate the
 elements of the three-dimensional representation. In section 4 we discuss its possible relevance to neutrino physics and present several  working examples.
 We summarize our results in section 5.

 \section{The  three-dimensional representations of the $PSL_2(7)$ group}

In the present section we describe the basic steps for the construction of  the unitary representations of the  $PSL_2(p)$ group.
This is  defined by the $2 \times 2$  matrices  $\mathfrak{a}$  with  elements  integers modulo $p$, where $p$ is a prime number, and
determinant equal to one modulo $p$:
\be
\mathfrak{a}=\left(\begin{array}{cc}a&b\\c&d\end{array}\right),\; a,b,c,d\in \mathbb{Z}_p,\;
{\rm det}\mathfrak{a}=1\,{\rm mod}{\,p}\label{2a2}
\ee
The elements of the group can be constructed from combinations of powers of  two generators denoted here with  $\mathfrak{a},
\mathfrak{b}$, which, in a specific representation  are defined by the matrices
\ba
\mathfrak{a}=\left(\begin{array}{cc}0&-1\\1&0\end{array}\right)\,;\;
\mathfrak{b}=\left(\begin{array}{cc}0&-1\\1&1\end{array}\right)\, \cdot \label{abgen}
\ea
These satisfy the relations
\be
\mathfrak{a}^2 =  \mathfrak{b}^3=-I\equiv - { \left(\begin{array}{cc}{ 1}&0\\0&{ 1}\end{array}\right)}\,\cdot
\label{abrel}
\ee
To define the projective linear group $PSL_2(p)$, we first observe that $SL_2(p)$ contains  a normal  subgroup   of two
elements, $Z_2=\{I, -I\}$.  The $PSL_2(p)$ is defined as the quotient subgroup, by identifying the unit matrix $I$ with $-I$
\be
PSL_2(p)= SL_2(p)/\{I, -I\}\;\cong \;SL_2(p)/Z_2
\ee
Next,  we use Weil's metaplectic representation { $U(A)$} derived long time ago  by Balian and Itzykson~\cite{BI},
( see also \cite{Athanasiu:1994fv}) to construct the $p$-dimensional unitary representations of  $SL_2(p)$ groups.
The explicit form of  $U(A)$ is given in terms of the elements $a,b,c,d$ of the $2\times 2$
matrix (\ref{2a2}) as follows
\small
\begin{eqnarray}
	&&{ U(A)}={\frac{ \sigma(1)\sigma({\delta})}{ p}}\sum_{ r,s}{\omega}^{[b{ r}^2+(d-a){ rs}-c{ s}^2]/(2{\delta})}{J_{ r,s}}
	\label{metap}
\end{eqnarray}
\normalsize
for ${ \delta}={ 2-a-d}\ne 0$.  For $\delta =0$, we distinguish the following cases:
\begin{eqnarray}
	{\delta } = 0,\; b\ne 0:&& U(A)=\frac{\sigma(-2b)}{\sqrt{p}}\sum_{s}\omega^{s^2/(2b)}{J_{s(a-1)/b,s}}
	\nonumber
	\\
	{ \delta} ={ b}=0,\; c\ne 0:&& U(A)=\frac{\sigma(2c)}{\sqrt{p}}\sum_{r}\omega^{-r^2/(2c)}{ P^r}
	\\
	{ \delta }= { b}=0= c= 0:&&{ U(1)=I}\nn
\end{eqnarray}
\normalsize
A few clarifications on notation and definitions in the above formulae  are needed.

\noindent
The quantities $J_{r,s}, J_{s(a-1)/b,s}$ are defined as follows
\be
J_{n_1,n_2}\equiv  J_{\vec n}=\omega^{\frac{n_1n_2}2} P^{n_1}Q^{n_2}~,\label{Jgen}
\ee
where $\omega = e^{2\pi i/p}$ is the $p^{th}$ root of unity,  while $P, Q$ are position and momentum operators with
elements  $P_{kl}=\delta_{k-1,l}$ and,  $ Q_{kl}=\omega^k\delta_{kl}$ repsectively.
The generators (\ref{Jgen}) obey the `multiplication'  law
\[{ J_{\vec m} J_{\vec n}}= { \omega}^{\frac{\vec n\times \vec m}2} { J_{\vec m+\vec n}}\]
and constitute a subset of the Heisenberg group~\cite{BI}.

\noindent
 The quantities  $\sigma(a)$ and $\left({\frac{a}{p}}\right)$, are the Quadratic Gauss Sum and the Legendre symbol respectively.  These are defined as follows:
\ba
{\sigma(a)} &=&\frac{1}{\sqrt{p}}\,\sum_{k=0}^{p-1}\omega^{ak^2} =\left({\frac{a}{p}}\right)\times \left\{\begin{array}{cc}1&
	{\rm for}\, p=4k+1\\i&{\rm for}\, p=4k-1\end{array}\right.\label{QGS}
\ea
and
\ba
\left({\frac{a}{p}}\right)&=&
\left\{\begin{array}{cc}
	0&{\rm if} { a}\; {\rm devides}\; { p}
	\\+1&{\rm if}\; { a} = {\cal QR}\; { p}
	\\-1&{\rm if}\; { a} \ne {\cal QR}\; {p}
\end{array}\right.\label{LS}
\ea
where $ {\cal QR}$ means Quadratic Residue~\footnote{An integer  $q$ is called
	Quadratic Residue (${\cal QR}$) \,iff \, $\exists\, x:\, x^2={ q}\,{\rm mod}\,{ p}$.}.
	
The  so obtained $p$-dimensional representation  decomposes into  two irreducible unitary representations of dimensions
$\frac{p+ 1}2$ and $\frac{p-1}2$. These are discrete subgroups of the unitary groups $SU(\frac{p\pm 1}2)$ and for $p=7$
we obtain the 3-dimensional representation of the discrete group $PSL_2(7)$ which  is  a subgroup of  $SU(3)$.  Smaller $p$
values result to $A_3$ and $A_5$ groups which have been extensively studied, while the next value  of $p=11$ results to $PSL_2(11)$
 which does not contain triplet representations in its decompositions $(p=11=6+5)$, therefore  it is not of our primary interest.

The $SL_2(p)$ group with $p$ prime has  $p (p^2-1)$ elements and the corresponding projective $PSL_2(p)$ contains half of them. Therefore,
the  $PSL_2(7)$ has $168$ elements and it is  a simple discrete subgroup of $SU(3)$. Using the method described above we  can
construct~\cite{Floratos:2015jvu} the three-dimensional representations of  $PSL_2(7)$, satisfying the conditions
\be
 \mathfrak{a}^2=\mathfrak{b}^3 =(\mathfrak{ab})^7= [\mathfrak{a},\mathfrak{b}]=1  \label{CRSL7}
\ee
where the `commutator' for the group elements is defined as usual:
$[\mathfrak{a},\mathfrak{b}]=\mathfrak{a}^{-1}\mathfrak{b}^{-1}\mathfrak{ab}$.
The generators of the 3-dimensional unitary representation of the $PSL_2(7)$ group, associated with ${ \mathfrak{\mathfrak{a}}}$ and
${ \mathfrak{\mathfrak{b}}}$  of (\ref{2a2})  can be written in terms of the
  $7^{th}$ root of unity  ${\eta}=e^{2\pi i/{ 7}}$, as follows:
	\ba
 { A_{[3]}}&=&{ \frac{i}{\sqrt{7}}
		\left(\begin{array}{ccc}
			{\eta}^2-{\eta}^5&{ \eta}^6-{ \eta}&{ \eta}^3-{ \eta}^4\\
			{ \eta}^6-{ \eta}&{\eta}^4-{ \eta}^3&{ \eta}^2-{ \eta}^5\\
			{\eta}^3-{ \eta}^4&{ \eta}^2-{ \eta}^5&{ \eta}-{ \eta}^6
		\end{array}\right)}\label{GenA}
	\ea
	and
	\ba
 { B_{[3]}}
	&=&{\frac{i}{\sqrt{7}}
		\left(\begin{array}{ccc}
			{ \eta}-{\eta}^4&{ \eta}^4-{ \eta}^6&{ \eta}^6-1\\
			{ \eta}^5-1&{ \eta}^2-{ \eta}&{ \eta}^5-{ \eta}\\
			{ \eta}^2-{ \eta}^3&1-{\eta}^3&{ \eta}^4-{ \eta}^2
		\end{array}\right)}\,.\label{GenB}
	\ea
It can be readily checked that these satisfy the required relations:
\be
{A_{[3]}}^2={B_{[3]}}^3=({A_{[3]}B_{[3]}})^7= [{A_{[3]},B_{[3]}}]^4=I~\cdot
\ee

\noindent
Implementing the above method,	we can  find  the $PSL_2(7)$ group elements, imposing the appropriate conditions
on products of powers of  	$2\times 2$ matrices  given in (\ref{2a2}) and then using the metaplectic representation
to build the 3-dimensional representations  of $PSL_2(7)$.
However, in this section we will follow a different approach which, in our opinion, reveals some new and very interesting
properties of the group elements.

For all practical purposes however, it suffices to take the two generators and 	
 explicitly  construct  all the  elements of the
 three-dimensional unitary representations of $PSL_2(7)$ using the GAP system for computational
 discrete algebra available in the web\cite{GAP}. After some algebra it can be shown that the two generators
 can be written as
\be
A_{[3]}=
\begin{bmatrix}
\rho_{1} & -\rho_{2} & -\rho_{3}\\
-\rho_{2} & \rho_{3} & \rho_{1}\\
-\rho_{3} & \rho_{1} & \rho_{2}%
\end{bmatrix}\,,
\;\;
B_{[3]}=
\begin{bmatrix}
\rho_{1}\eta^{-1} & \rho_{2}\eta^{\frac{3}{2}} & \rho_{3}\eta^{-\frac{1}{2}}\\
\rho_{2}\eta^{\frac{5}{2}} & \rho_{3}\eta^{-2} & \rho_{1}\eta^{3}\\
\rho_{3}\eta^{\frac{5}{2}} & \rho_{1}\eta^{-2} & \rho_{2}\eta^{3}%
\end{bmatrix}
~\cdot
\ee
where
\be
\rho_{1}=-\frac{2}{\sqrt{7}}\cos\frac{\pi}{14},\;
\rho_{2}=-\frac{2}{\sqrt{7}}\cos\frac{3\pi}{14},\;
\rho_{3}=-1-\rho_{1}-\rho_{2}~\cdot\label{rhos}
\ee
The quantities $\rho_{1},\rho_{2},\rho_{3}$ satisfy the cubic equation
\[
x^{3}+x^{2}-\frac{1}{7}=0~,
\]
and the relation $\rho_{2}=7\rho_{1}^{3}-3\rho_{1}~$.

\noindent
We now  observe that the moduli of the elements of both matrices follow a latin square pattern~\cite{latin}.
The so found group matrices can be classified according to their conjugacy class as shown in Table~\ref{ConClass}.
\begin{table}
	\begin{center}
\begin{tabular}
{cclc}
Order & character & $\#$ & Tag\\\hline
${2}$ & $-1$ & $21$ & $el_{2}$\\
${3}$ & $\hspace{0.275cm} 0$ & $56$ & $el_{3}$\\
${4}$ & $+1$ & $42$ & $el_{4}$\\
${7}$ & $-\frac{1}{2}\pm i\frac{\sqrt{7}}{2}$ & $48$ & $el_{7}$\\
\end{tabular}
\caption{The order, character, and the number of elements of the conjugacy classes of $PSL_2(7)$}
\end{center}\label{ConClass}
\end{table}

\bigskip

\section{On the properties of the representation matrices}

  In this section we investigate useful  properties  of  latin square matrices in view of  their relation to the
  $PSL_2(7)$ elements.  We start with the $3\times3$ case with real entries. These are of the following two types
 \begin{equation}
 M_{1}=
 \begin{bmatrix}
 r_{1} & r_{2} & r_{3}\\
 r_{2} & r_{3} & r_{1}\\
 r_{3} & r_{1} & r_{2}%
 \end{bmatrix},\;
 M_{2}=%
 \begin{bmatrix}
 r_{1} & r_{2} & r_{3}\\
 r_{3} & r_{1} & r_{2}\\
 r_{2} & r_{3} & r_{1}%
 \end{bmatrix}\,,
 \end{equation}
 and their permutations. From these two, only the  first type appears in  $PSL_2\left(7\right)$.
 The orthogonality condition  $M_{1}^{2}=1$  implies that
\begin{equation}
r_{1}^{2}+r_{2}^{2}+r_{3}^{2}=1,\;
r_{1}r_{2}+r_{1}r_{3}+r_{2}r_{3}=0\label{symr}
\end{equation}
while requiring  $\det M_{1}=1$ we get
\begin{equation}
r_{1}+r_{2}+r_{3}=-1\label{sumr}\cdot
\end{equation}
Thus, if the matrix $M_1$ is part of an irreducible group representation,  it should belong to the conjugacy class
$el_2$ with character $-1$. Notice that  $r_{1},r_{2}, r_{3}$ satisfy the algebraic equation
\begin{equation}
x^3+x^2-q=0,\;  {\rm where } \; q= r_1r_2r_3\label{qeq} ,
\end{equation}
and the  reality of the roots requires that ${\small 0< q<\frac{4}{27}}\cdot $ For the case of $PSL_2(7)$,
$q=\frac 17<\frac{4}{27} \cdot$

The obvious generalization includes complex elements and takes the form
\begin{equation}
M=%
\begin{bmatrix}
r_{1}e^{\imath c_{1}} & r_{2}e^{\imath c_{2}} & r_{3}e^{\imath c_{3}}\\
r_{2}e^{\imath c_{4}} & r_{3}e^{\imath c_{5}} & r_{1}e^{\imath c_{6}}\\
r_{3}e^{\imath c_{7}} & r_{1}e^{\imath c_{8}} & r_{2}e^{\imath c_{9}}%
\end{bmatrix}
~.
\end{equation}
Unitarity and  the condition det$M=1$ restrict the number of free parameters $r_i, c_i$.
The conditions (\ref{symr},\ref{sumr}) still hold, and  the final form is
\begin{equation}
M=%
\begin{bmatrix}
r_{1}e^{ic_{1}} & r_{2}e^{ic_{2}} & r_{3}e^{ic_{3}}\\
r_{2}e^{i\left(  c_{1}-c_{2}+c_{5}\right)  } & r_{3}e^{ic_{5}} &
r_{1}e^{i\left(  c_{3}-c_{2}+c_{5}\right)  }\\
r_{3}e^{-i\left(  c_{3}+c_{5}\right)  } & r_{1}e^{i\left(  c_{2}-c_{3}%
-c_{1}-c_{5}\right)  } & r_{2}e^{-i\left(  c_{1}+c_{5}\right)  }%
\end{bmatrix}\label{cm} \cdot
\end{equation}
Looking more closely at the $PSL_{2}(7)$ elements constructed by GAP, we observe  that a great number of them can be written in this form
when we substitute $\rho_{1},\rho_{2},\rho_{3}$ given in (\ref{rhos}) in  place of $r_{1},r_{2},r_{3}$  and all the phases are closely
related to the set  of the seventh roots of unity.  The secular equation for $M$ reads
\begin{equation}
x^{3}-tr\left(  M\right)  x^{2}+trM^{\ast}x-1=0~\cdot\label{seceq}
\end{equation}
It can be readily checked that given the character of the conjugacy class  Tr$\left(  M\right)$, the secular
 equation  (\ref{seceq}) reproduces the correct eigenvalues  of the representation matrices.
 For example, the order seven elements character is  Tr$M=-\frac{1}{2}-i\frac{\sqrt{7}}{2}$ and
solving the secular equation we find the eigenvalues
\begin{equation}
\exp\left[  \frac{10\pi i}{7}\right],\;\exp\left[  \frac{6\pi i}{7}\right],\;\exp\left[  \frac{12\pi i}{7}\right],
\end{equation}
in accordance to the group elements table. For the order two elements a
trivial calculation implies that $c_{1}=c_{5}=0$ giving a general form
\begin{equation}
M=%
\begin{bmatrix}
r_{1} & r_{2}e^{\imath c_{2}} & r_{3}e^{\imath c_{3}}\\
r_{2}e^{-\imath c_{2}} & r_{3} & r_{1}e^{\imath\left(  c_{3}-c_{2}\right)  }\\
r_{3}e^{-\imath c_{3}} & r_{1}e^{\imath\left(  c_{2}-c_{3}\right)  } & r_{2}%
\end{bmatrix}
~\cdot
\end{equation}

Next, for the general matrix $M$, information on the allowed values for $c_1$ and $c_5$ can be extracted by taking the system of equations
\ba
r_1 e^{ic_1}+r_2 e^{-i(c_1+c_5)}+r_3 e^{i c_5}&=&{\rm Tr} M\label{TrM}\\
r_1+r_2+r_3&=&-1\\
r_1r_2+r_2r_3+r_3r_1&=&0\label{TrM3}
\ea
and substituting the character ${\rm Tr} M$  of the corresponding conjugacy class.
Parametrizing the phases as
\be
c_{1}=\frac{2\pi}{7}n,\;  c_{5}=\frac{2\pi}{7}m\label{cmn}
\ee
and  taking $q=r_1r_2r_3=\frac 17$ we get  only integer values for $n, m$.
  These values are symmetric  under the interchange of  $m$ and $n$.
Notice that the  value, $q=\frac{1}{7}$ identifies
$r_{1}\allowbreak, r_{2}, r_{3}$ with $\rho_{1},\rho_{2},\rho_{3}$
respectively.

Solving the system of  equations~(\ref{TrM})-(\ref{TrM3}), for the order $3$ elements we have
\begin{equation}%
\begin{bmatrix}
r_{1}\\
r_{2}\\
r_{3}%
\end{bmatrix}
=%
\begin{bmatrix}
1 & 1 & 1\\
\cos c_{1} & \cos\left(  c_{1}+c_{5}\right)   & \cos c_{5}\\
\sin c_{1} & -\sin\left(  c_{1}+c_{5}\right)   & \sin c_{5}%
\end{bmatrix}
^{-1}%
\begin{bmatrix}
-1\\
0\\
0
\end{bmatrix}
~,
\end{equation}
and the solutions found are
\[%
\begin{tabular}
[c]{rrrrrrrr}
$\mathbf{c}_{1}$ & $\mathbf{c}_{5}$ & $\mathbf{c}_{1}$ & $\mathbf{c}_{5}$ &
$\mathbf{c}_{1}$ & $\mathbf{c}_{5}$ & $\mathbf{c}_{1}$ & $\mathbf{c}_{5}%
$\\
$\mathbf{n}$ & $\mathbf{m}$ & $\mathbf{n}$ & $\mathbf{m}$ & $\mathbf{n}$ &
$\mathbf{m}$ & $\mathbf{n}$ & $\mathbf{m}$\\
\hline
$-5$ & $-3$ & $-4$ & $-2$ & $-2$ & $-1$ & $1$ & $4$\\
$-5$ & $1$ & $-4$ & $5$ & $-2$ & $3$ & $1$ & $2$\\
$-5$ & $4$ & $-3$ & $1$ & $-1$ & $3$ & $2$ & $4$\\
$-4$ & $-1$ & $-3$ & $2$ & $-1$ & $5$ & $3$ & $5$\\
\end{tabular}
\ ~\cdot
\]
For the order 4  elements we have
\begin{equation}%
\begin{bmatrix}
r_{1}\\
r_{2}\\
r_{3}%
\end{bmatrix}
=%
\begin{bmatrix}
1 & 1 & 1\\
\cos c_{1} & \cos\left(  c_{1}+c_{5}\right)   & \cos c_{5}\\
\sin c_{1} & -\sin\left(  c_{1}+c_{5}\right)   & \sin c_{5}%
\end{bmatrix}
^{-1}%
\begin{bmatrix}
-1\\
1\\
0
\end{bmatrix}
~,
\end{equation}
which imply the following values for $m,n$ and $c_{1,5}$
\[
\begin{tabular}
[c]{rrrrrrrr}\hline
$\mathbf{c}_{1}$ & $\mathbf{c}_{5}$ & $\mathbf{c}_{1}$ & $\mathbf{c}_{5}$ &
$\mathbf{c}_{1}$ & $\mathbf{c}_{5}$ & $\mathbf{c}_{1}$ & $\mathbf{c}_{5}$ \\
$\mathbf{n}$ & $\mathbf{m}$ & $\mathbf{n}$ & $\mathbf{m}$ & $\mathbf{n}$ &
$\mathbf{m}$ & $\mathbf{n}$ & $\mathbf{m}$
\\\hline
$-6$ & $-3$ & $-5$ & $1$ & $-4$ & $5$ & $-2$ & $6$ \\
$2$ & $4$&$-6$ & $-5$ & $-5$ & $4$ & $-3$ & $1$\\
$-1$ & $3$ & $3$ & $6$&$-6$ & $4$ & $-4$ & $-1$ \\
$-3$ & $2$ & $-1$ & $5$ & $3$ & $5$&$-6$ & $2$ \\
 $-4$ & $-2$ & $-2$ & $-1$ & $1$ & $4$ & $5$ & $6$\\
$-5$ & $-3$ & $-4$ & $6$ & $-2$ & $3$ & $1$ & $2$  \\
\end{tabular}
\]
For the order $7$ elements we have
\begin{equation}%
\begin{bmatrix}
r_{1}\\
r_{2}\\
r_{3}%
\end{bmatrix}
=%
\begin{bmatrix}
1 & 1 & 1\\
\cos c_{1} & \cos\left(  c_{1}+c_{5}\right)   & \cos c_{5}\\
\sin c_{1} & -\sin\left(  c_{1}+c_{5}\right)   & \sin c_{5}%
\end{bmatrix}
^{-1}%
\begin{bmatrix}
-1\\
-\frac{1}{2}\\
-\frac{\sqrt{7}}{2}%
\end{bmatrix}
~\cdot
\end{equation}
The solutions are%
\[%
\begin{tabular}
[c]{rrrrrr}
$\mathbf{c}_{1}$ & $\mathbf{c}_{5}$ & $\mathbf{c}_{1}$ & $\mathbf{c}_{5}$ &
$\mathbf{c}_{1}$ & $\mathbf{c}_{5}$\\\hline
$\mathbf{n}$ & $\mathbf{m}$ & $\mathbf{n}$ & $\mathbf{m}$ & $\mathbf{n}$ &
$\mathbf{m}$\\\hline
$-6$ & $-3$ & $-5$ & $-3$ & $-3$ & $2$\\
$-6$ & $-5$ & $-5$ & $1$ & $1$ & $4$\\
$-6$ & $4$ & $-5$ & $4$ & $1$ & $2$\\
$-6$ & $2$ & $-3$ & $1$ & $2$ & $4$\\
\end{tabular}
\]
Note that all these solutions when applied to the system for $r_{1}, r_{2}, r_{3}$  just generate permutations of  the root system $\rho_{1},
 \rho_{2}, \rho_{3}$. Concerning the possible values of the remaining phases $c_{2}$ and $c_{3}$ no further information can be extracted without
using the group algebra.

\section{$SL_2(7)$ invariance and the neutrino mixing matrix}

Neutrino oscillations are  associated with the existence of non-zero  masses $m_{\nu_i}$ and  non-zero
$\theta_{ij}$ mixing angles in the lepton sector.  In a standard parametrization the corresponding lepton mixing matrix
 is given by the expression
\begin{equation}
U=U_{l}^{\dagger }U_{\nu }= \left(
\begin{array}
[c]{lll}%
c_{12}c_{13} & c_{13}s_{12} & s_{13}e^{-i\delta}\\
-c_{23}s_{12}-c_{12}s_{13}s_{23}e^{i\delta} & c_{12}c_{23}-s_{12}s_{13}s_{23}e^{i\delta} &
c_{13}s_{23}\\
s_{12}s_{23}-c_{12}c_{23}s_{13}e^{i\delta} & -c_{23}s_{12}s_{13}-c_{12}s_{23}e^{i\delta} & c_{13}c_{23}%
\end{array}
\right) \label{V123}%
\end{equation}
where, to avoid clutter, we have denoted $c_{ij}\equiv\cos\theta_{ij}$ and $s_{ij}\equiv\sin\theta_{ij}$.
The $3\sigma$ range of the three mixing angles in accordance with recent data, is given by
\be
	\sin^{2}\theta_{12}  =[0.25-0.35],\;
	\sin^{2}\theta_{23}  =[0.38-0.62],\;
	\sin^{2}\theta_{13}  =[0.0185-0.0246]\cdot
\ee
We will confront our results on the masses and mixing matrices with the experimental data.

\noindent
In order to construct the mixing matrices one has to first assume the
symmetries of the neutrino and charged leptons mass matrices. These
symmetries are connected to the elements of $PSL_{2}\left( 7\right) $ which
leave the mass matrices invariant (i.e. vanishing commutator). Then we
determine the diagonalizing matrices $U$ for these elements.
Note that the diagonalizing matrices are not uniquely defined since there
are $3!$ ways to arrange the eigenvalues in the resulting diagonal matrices.
Since $PSL_{2}\left( 7\right) $ contains four conjugacy classes characterized
by elements of order $2$ $\left( el_{2}\right) $, order $3$ $\left(
el_{3}\right) $, order $4$ $\left( el_{4}\right) $, and $7$ $\left(
el_{7}\right) $ (see appendix for notation),  in order to construct the mixing matrices one has to combine
the diagonalizing matrices in all possible ways. This search of course can
only be done numerically. In order to conform with experimental data we kept
only the cases where $0.136<\left\vert U_{13}\right\vert <0.157$, $%
0.499<\left\vert U_{12}\right\vert <0.595$, $0.615<\left\vert
U_{23}\right\vert <0.785$.
It turns out that the only symmetry for the
neutrino mass matrix that is compatible with data is connected to a number
of order $2$ elements. For the charged leptons mass matrix the symmetry
allowed is connected to both order $3$ and order $7$ elements. The results
are shown in the tables~\ref{tel3} and~\ref{tel7}. Obviously, these tables should also
contain the inverse elements. Since the order $2$ elements equal their
inverses and the rest can be easily calculated, the inverses are not shown
for reasons of  clarity.
\begin{table}
	\begin{center}
\begin{tabular}{lllllllll}
$\mathbf{el}_{2}$ & $\mathbf{el}_{3}$ & $\mathbf{el}_{3}$ & $\mathbf{el}_{3}$
& $\mathbf{el}_{3}$ & $\mathbf{el}_{3}$ & $\mathbf{el}_{3}$ & $\mathbf{el}%
_{3}$ & $\mathbf{el}_{3}$ \\
\hline
$15$ & \multicolumn{1}{c}{$6$} & \multicolumn{1}{c}{$7$} &
\multicolumn{1}{c}{$15$} & \multicolumn{1}{c}{$17$} & \multicolumn{1}{c}{$23$%
} & \multicolumn{1}{c}{$25$} & \multicolumn{1}{c}{$37$} & \multicolumn{1}{c}{%
$41$} \\
$16$ & \multicolumn{1}{c}{$5$} & \multicolumn{1}{c}{$6$} &
\multicolumn{1}{c}{$19$} & \multicolumn{1}{c}{$21$} & \multicolumn{1}{c}{$24$%
} & \multicolumn{1}{c}{$26$} & \multicolumn{1}{c}{$39$} & \multicolumn{1}{c}{%
$42$} \\
$17$ & \multicolumn{1}{c}{$4$} & \multicolumn{1}{c}{$5$} &
\multicolumn{1}{c}{$16$} & \multicolumn{1}{c}{$18$} & \multicolumn{1}{c}{$25$%
} & \multicolumn{1}{c}{$27$} & \multicolumn{1}{c}{$38$} & \multicolumn{1}{c}{%
$41$} \\
$18$ & \multicolumn{1}{c}{$3$} & \multicolumn{1}{c}{$4$} &
\multicolumn{1}{c}{$15$} & \multicolumn{1}{c}{$20$} & \multicolumn{1}{c}{$26$%
} & \multicolumn{1}{c}{$28$} & \multicolumn{1}{c}{$40$} & \multicolumn{1}{c}{%
$42$} \\
$19$ & \multicolumn{1}{c}{$1$} & \multicolumn{1}{c}{$3$} &
\multicolumn{1}{c}{$17$} & \multicolumn{1}{c}{$19$} & \multicolumn{1}{c}{$22$%
} & \multicolumn{1}{c}{$27$} & \multicolumn{1}{c}{$36$} & \multicolumn{1}{c}{%
$38$} \\
$20$ & \multicolumn{1}{c}{$1$} & \multicolumn{1}{c}{$8$} &
\multicolumn{1}{c}{$16$} & \multicolumn{1}{c}{$21$} & \multicolumn{1}{c}{$23$%
} & \multicolumn{1}{c}{$28$} & \multicolumn{1}{c}{$37$} & \multicolumn{1}{c}{%
$40$} \\
$21$ & \multicolumn{1}{c}{$7$} & \multicolumn{1}{c}{$8$} &
\multicolumn{1}{c}{$18$} & \multicolumn{1}{c}{$20$} & \multicolumn{1}{c}{$22$%
} & \multicolumn{1}{c}{$24$} & \multicolumn{1}{c}{$36$} & \multicolumn{1}{c}{%
$39$}\\
\end{tabular}
\caption{Solutions for the charged lepton mass matrix in terms of the order 3 elements.}
\end{center}\label{tel3}
\end{table}

\begin{table}
	\begin{center}
\begin{tabular}{lllllllll}
$\mathbf{el}_{2}$ & $\mathbf{el}_{7}$ & $\mathbf{el}_{7}$ & $\mathbf{el}_{7}$
& $\mathbf{el}_{7}$ & $\mathbf{el}_{7}$ & $\mathbf{el}_{7}$ \\
\hline
$8$ & $3$ & $5$ & $15$ & $16$ & $21$ & $24$ \\
$9$ & $1$ & $2$ & $17$ & $18$ & $22$ & $25$ \\
$10$ & $4$ & $6$ & $12$ & $13$ & $23$ & $26$ \\
$11$ & $3$ & $7$ & $14$ & $15$ & $20$ & $24$ \\
$12$ & $1$ & $5$ & $16$ & $17$ & $21$ & $25$ \\
$13$ & $2$ & $4$ & $12$ & $18$ & $22$ & $26$ \\
$14$ & $6$ & $7$ & $13$ & $14$ & $20$ & $23$%
\end{tabular}
\caption{Solutions for the charged lepton mass matrix in terms of the order 7 elements.}
\end{center}\label{tel7}
\end{table}

Some comments concerning the order $2$ elements are here in order. The
eigenvalues of these matrices are $\left( 1,-1,-1\right) $ respectively,
i.e. there exists a degenerate 2-dimensional subspace implying that the eigenvectors
related to the degenerate eigenvalue cannot be uniquely defined. In fact, if
$v_{1},v_{2},v_{3}$ are eigenvectors corresponding to the $\left(
1,-1,-1\right) $ eigenvalues an equally good choice would be $v_{1},\widetilde{%
v_{2}},\widetilde{v_{3}}$ where
\be
\begin{bmatrix}
\widetilde{v_{2}} \\
\widetilde{v_{3}}%
\end{bmatrix}%
=%
\begin{bmatrix}
e^{i\varphi _{1}}\cos \varphi  & -e^{i\varphi _{2}}\sin \varphi \\
e^{i\varphi _{1}}\sin \varphi  & e^{i\varphi _{2}}\cos \varphi
\end{bmatrix}%
\begin{bmatrix}
v_{2} \\
v_{3}%
\end{bmatrix}~\cdot \label{U2rot}
\ee
for arbitrary $\varphi, \varphi_1, \varphi_2$.
This matrix defines a $U\left( 2\right) ~$rotation for $v_{2}$ and $v_{3}$
that leaves the representation matrices invariant.

\subsection{Working Examples}

In the following we analyse a few working examples and show how to explicitly construct
the mixing matrices.

\subsubsection{The pair  $el_{2}\left( 16\right),\; el_{3}\left( 5\right) $}
The corresponding matrices are given by
\begin{equation}
el_{2}\left( 16\right) =%
\begin{bmatrix}
r_{3} & -r_{1} & -r_{2} \\
-r_{1} & r_{2} & r_{3} \\
-r_{2} & r_{3} & r_{1}%
\end{bmatrix},\;
el_{3}\left( 5\right) =%
\begin{bmatrix}
0 & 0 & -e^{\frac{6\pi i}{7}} \\
e^{-\frac{2\pi i}{7}} & 0 & 0 \\
0 & e^{-\frac{4\pi i}{7}} & 0%
\end{bmatrix}%
~\cdot
\end{equation}%
The normalized eigenvectors of $el_{2}\left( 16\right)$ are given by
\[
{  v_{2}\left[ 1\right] =%
\begin{bmatrix}
s \\
\frac{1}{2}s-\frac{\sqrt{3}}{2}\sqrt{\frac{2}{3}-s^{2}} \\
\frac{1}{2}s+\frac{\sqrt{3}}{2}\sqrt{\frac{2}{3}-s^{2}}%
\end{bmatrix}},
\;
{  v_{2}\left[ 2\right] =\frac{1}{\sqrt{3}}%
\begin{bmatrix}
-1 \\
1 \\
1%
\end{bmatrix}},
\;
{v_{2}\left[ 3\right] =%
\begin{bmatrix}
\sqrt{\frac{2}{3}-s^{2}} \\
\frac{\sqrt{3}}{2}s+\frac{1}{2}\sqrt{\frac{2}{3}-s^{2}} \\
-\frac{\sqrt{3}}{2}s+\frac{1}{2}\sqrt{\frac{2}{3}-s^{2}}%
\end{bmatrix}}
\]%
related to the eigenvalues $+1$, $-1,$ $-1$ respectively, with $s$ given by
\be
s=\sqrt{2}\frac{\left( r_{1}-r_{2}\right) }{\sqrt{\left( 1-3r_{3}\right)
^{2}+3\left( r_{2}-r_{1}\right) ^{2}}}\approx -0.815~~\cdot
\label{sval}
\ee

\begin{figure}[t]
	\begin{center}
		\includegraphics[width=0.465\textwidth,angle=0,scale=0.9]{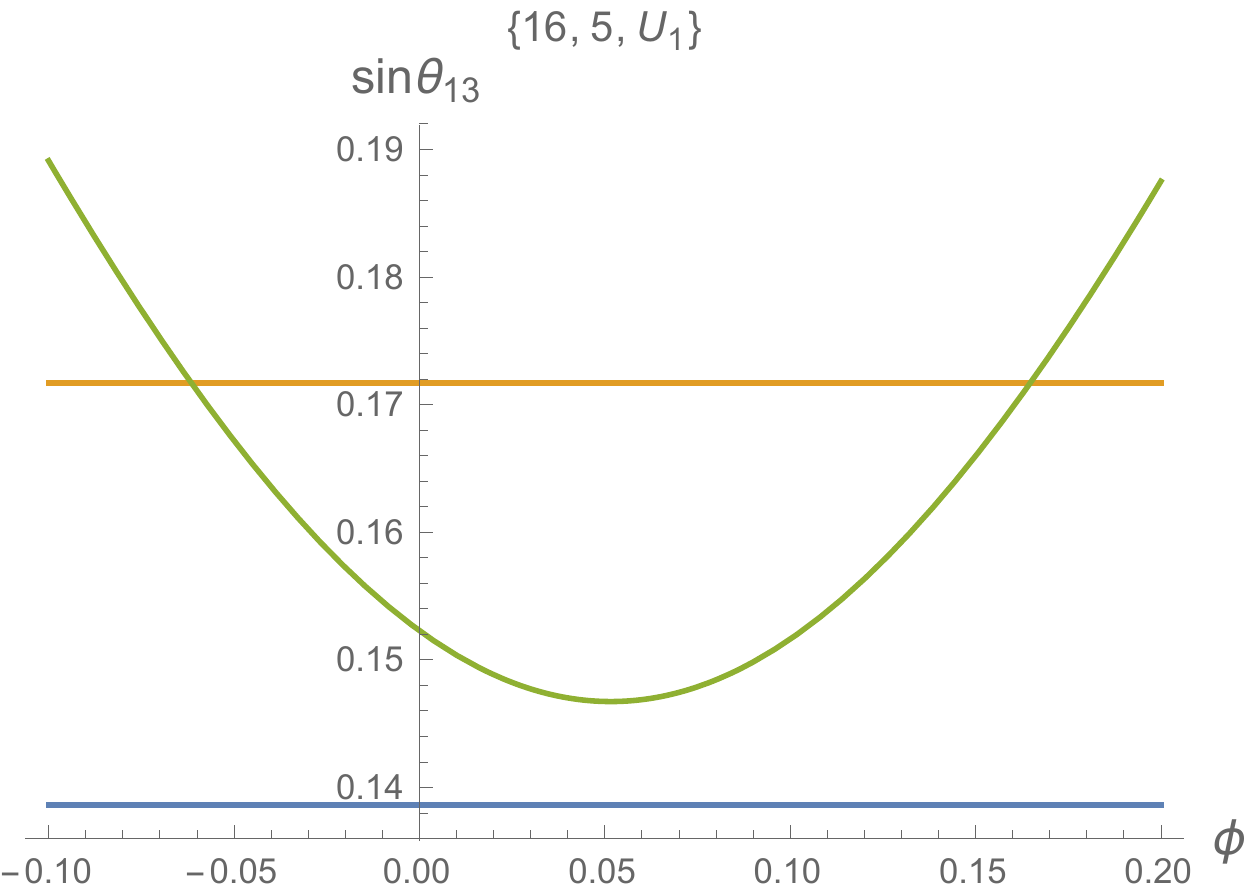}
		\hspace{0.1cm}
		\includegraphics[width=0.465\textwidth,angle=0,scale=0.9]{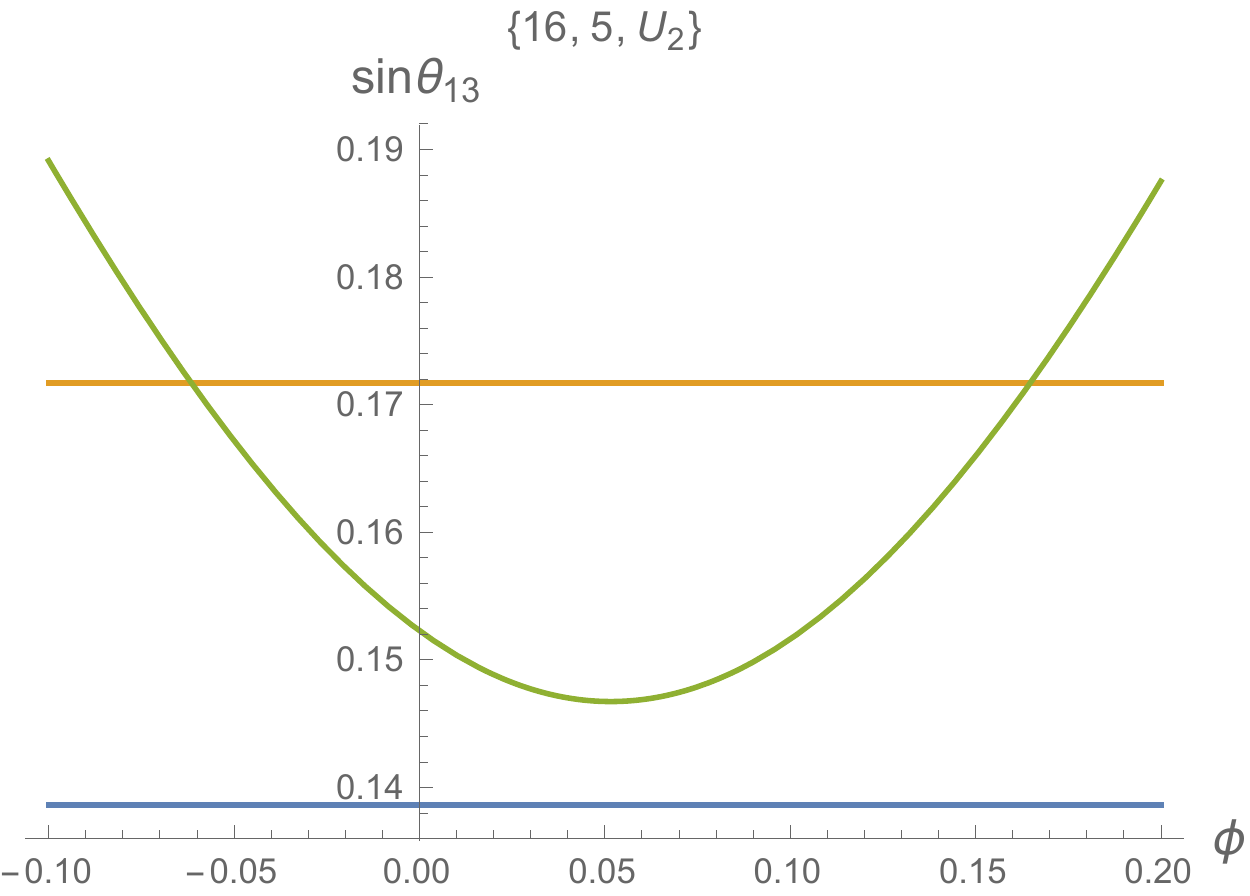}
	\end{center}
	\caption{
	Case  $el_{2}\left( 16\right),\; el_{3}\left( 5\right) $:	The range of $\sin\theta_{13}$ as a function of the angle $\phi$ parametrizing the
		mixing ~(\ref{U2rot}) of the degenerate subspace. The orange line defines the
upper experimental bound and the blue the lower one on $\sin\theta_{13}$.	}
	\label{fig:1}
\end{figure}

The normalized eigenvectors of $el_{3}(5)$ are
\begin{equation}
v_{3}\left[ 1\right] =\frac{1}{\sqrt{3}}
\begin{bmatrix}
-e^{\frac{6\pi i}{7}} \\
e^{\frac{4\pi i}{7}} \\
1
\end{bmatrix}, v_{3}\left[ 2\right] =\frac{1}{\sqrt{3}}%
\begin{bmatrix}
-e^{\frac{4\pi i}{21}} \\
-e^{\frac{5\pi i}{21}} \\
1
\end{bmatrix}, v_{3}\left[ 3\right] =\frac{1}{\sqrt{3}}%
\begin{bmatrix}
e^{\frac{11\pi i}{21}} \\
-e^{\frac{19\pi i}{21}} \\
1
\end{bmatrix}%
\end{equation}
related to the eigenvalues $1$, $e^{\frac{2\pi i}{3}}$, $e^{\frac{4\pi i}{3}}$
correspondingly.   Further manipulation shows that a compatible with data mixing
matrix occurs only for the following two combinations%
\ba
U_{1}&=&
\begin{bmatrix}
v_{3}\left[  3\right],   & v_{3}\left[  2\right],   & v_{3}\left[  1\right]
\end{bmatrix}
^{\dagger}\cdot
\begin{bmatrix}
\tilde v_{2}\left[  2\right],   & v_{2}\left[  1\right],   &\tilde  v_{2}\left[  3\right]
\end{bmatrix}
\nn\\
U_{2}&=&
\begin{bmatrix}
v_{3}\left[  3\right],   & v_{3}\left[  1\right],   & v_{3}\left[  2\right]
\end{bmatrix}
^{\dagger}\cdot
\begin{bmatrix}
\tilde v_{2}\left[  2\right],   & v_{2}\left[  1\right],   &\tilde  v_{2}\left[  3\right]
\end{bmatrix}
\ea
where $\tilde v_2[2], \tilde v_2[3], $ as in~(\ref{U2rot}).
Explicit calculations show that the modulus of the second column elements for
both matrices $U_{1}$ and  $U_{2}$ is $1/\sqrt{3}$. This is due to the fact
that
\begin{equation}
\left\Vert v_{3}\left[  2\right]  ^{\dagger}\cdot v_{2}\left[  1\right]  \right\Vert
=\left\Vert v_{3}\left[  1\right]  ^{\dagger}\cdot v_{2}\left[  1\right]
\right\Vert =\left\Vert v_{3}\left[  3\right]  ^{\dagger}\cdot v_{2}\left[
1\right]  \right\Vert =\frac{1}{\sqrt{3}}\approx 0.5773
\end{equation}
only for the given value of $s$~(\ref{sval}) and, surprisingly, is not obviously related to
the existence of the degenerate subspace.
For the pair $el_{2}\left(  16\right)  $ and $el_{3}\left(  5\right)  $ the
mixing matrices for $\varphi=0$ are given by%
\ba
U_{1}&=&
\begin{bmatrix}
0.80217e^{0.5667i} & 0.57735e^{2.3948i} & 0.152283e^{-1.27039i}\\
0.36647e^{0.106487i} & 0.57735e^{-0.8735i} & 0.729634e^{-0.3499i}\\
0.471405e^{-1.6582i} & 0.57735e^{3.05416i} & 0.666667e^{0.635302i}%
\end{bmatrix}
\\
U_{2}&=&
\begin{bmatrix}
0.80217e^{0.5667i} & 0.57735e^{2.3948i} & 0.152283e^{-1.27039i}\\
0.471405e^{-1.6582i} & 0.57735e^{3.05416i} & 0.666667e^{0.635302i}\\
0.36647e^{0.106487i} & 0.57735e^{-0.8735i} & 0.729634e^{-0.3499i}%
\end{bmatrix}
~\cdot\label{mixnum}
\ea
More generally, using  the degeneracy of the  subspace ~(\ref{U2rot}) we can  determine the range of $\varphi$  in accordance with the
experimental data. This is depicted in figure~\ref{fig:1}.
 \begin{figure}[t]
 	\begin{center}
 		\includegraphics[width=0.465\textwidth,angle=0,scale=0.9]{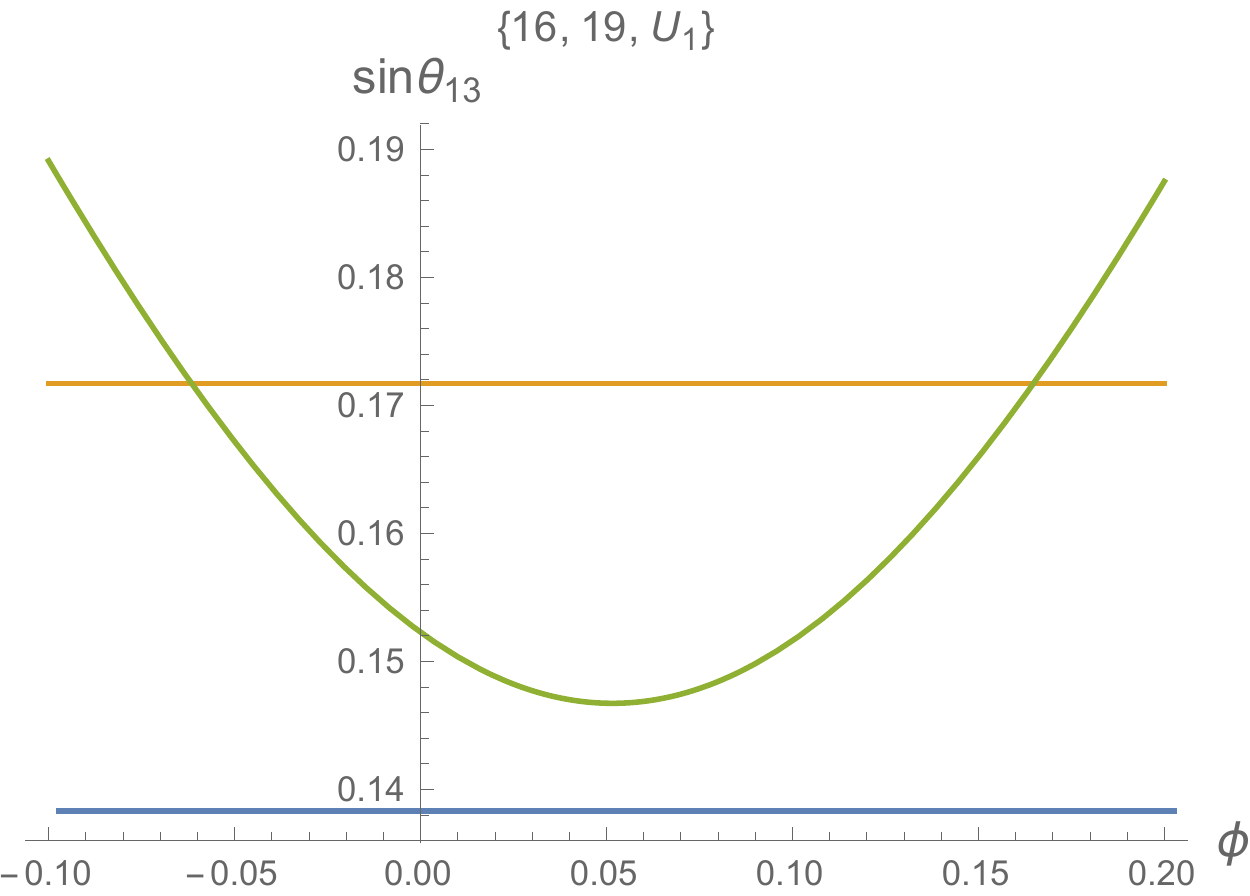}
 		\hspace{0.1cm}
 		\includegraphics[width=0.465\textwidth,angle=0,scale=0.9]{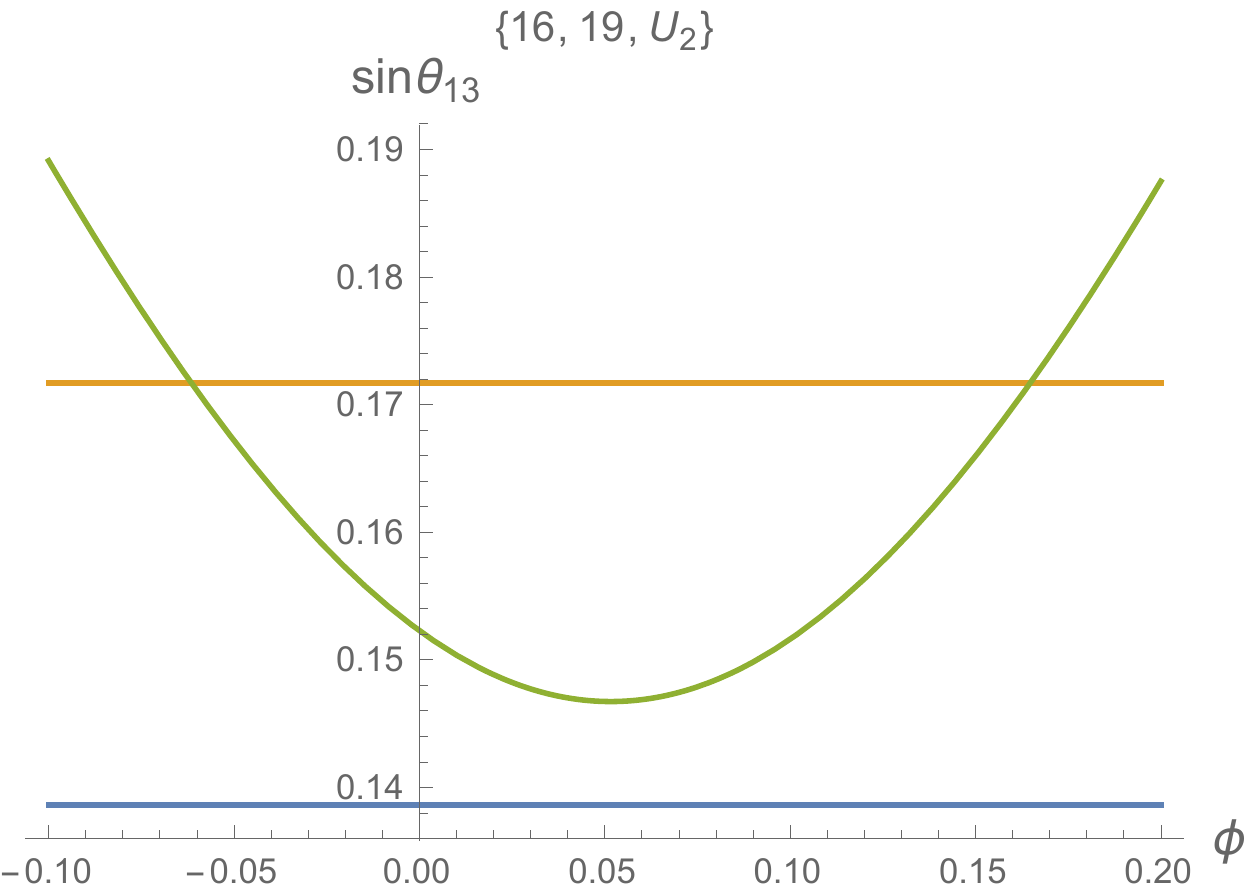}
 	\end{center}
 	\caption{
 		The acceptable range of $\sin\theta_{13}$ vs the angle $\varphi$ for the pair
 	 $el_{2}\left(  16\right)  ~el_{3}\left(  19\right)  $ .}
 	\label{fig:2}
 \end{figure}

\subsubsection{The pair $el_{2}\left(  16\right)  ~el_{3}\left(  19\right)  $ }
The relevant matrix is
\begin{equation}
el_{3}\left(  19\right)  =%
\begin{bmatrix}
r_{1}e^{\frac{2\pi i}{7}} & r_{2}e^{-\frac{5\pi i}{7}} & r_{3}e^{-\frac{5\pi
i}{7}}\\
r_{2}e^{-\frac{3\pi i}{7}} & r_{3}e^{\frac{4\pi i}{7}} & r_{1}e^{\frac{4\pi
i}{7}}\\
r_{3}e^{\frac{\pi i}{7}} & r_{1}e^{-\frac{6\pi i}{7}} & r_{2}e^{-\frac{6\pi
i}{7}}%
\end{bmatrix}
~.
\end{equation}
The  eigenvectors  which correspond to the eigenvalues
$1$, $e^{\frac{2\pi i}{3}}$, $e^{\frac{4\pi i}{3}}$ are
\ba
v_{3}\left[  1\right] &=&N_1\{
r_{2}\left( \eta ^{-\frac{5}{2}}+\eta ^{\frac{1}{2}}\right) ,  \eta
^{3}-r_{1}\eta ^{4}+r_{3}\eta -r_{2},  r_{1}\left( 1+\eta \right)\}\\
v_{3}\left[
2\right]&=&N_2\{
r_{2}\left( \eta ^{-\frac{5}{2}}+\eta ^{\frac{17}{6}}\right) ,  \eta
^{\frac{23}{3}}-\eta ^{\frac{7}{3}}\left( r_{2}+r_{1}\eta ^{4}\right)
+r_{3}\eta , r_{1}\left( \eta ^{\frac{7}{3}}+\eta \right)\}\\
v_{3}\left[  3\right]&=&N_3\left\{
r_{2}\left( \eta ^{-\frac{5}{2}}+\eta ^{\frac{31}{6}}\right) ,   \eta
^{\frac{37}{3}}-\eta ^{\frac{14}{3}}\left( r_{2}+r_{1}\eta ^{4}
\right) +r_{3}\eta ,  r_{1}\left( \eta ^{\frac{14}3}+\eta \right)\right\}
\ea
where $N_1$ $N_2$ and $N_3$ are normalization factors.
We find that a mixing matrix compatible  with data occurs only for the two combinations%
\ba
U_{1}&=&
\begin{bmatrix}
v_{3}\left[  2\right],   & v_{3}\left[  1\right],  & v_{3}\left[  3\right]
\end{bmatrix}
^{\dagger}\cdot
\begin{bmatrix}
\tilde v_{2}\left[  3\right],   & v_{2}\left[  1\right],   &\tilde  v_{2}\left[  2\right]
\end{bmatrix}
\\
U_{2}&=&
\begin{bmatrix}
v_{3}\left[  2\right],   & v_{3}\left[  3\right],   & v_{3}\left[  1\right]
\end{bmatrix}
^{\dagger}\cdot
\begin{bmatrix}
\tilde v_{2}\left[  3\right],   & v_{2}\left[  1\right],   & \tilde v_{2}\left[  2\right]
\end{bmatrix}
~\cdot
\ea
In this case also, we find that the modulus of the second column elements for
both matrices $U_{1}$and $U_{2}$ is $1/\sqrt{3}$. This is due to the fact
that
\begin{equation}
\left\Vert v_{3}\left[  1\right]  ^{\dagger}\cdot v_{2}\left[  1\right]  \right\Vert
=\left\Vert v_{3}\left[  2\right]  ^{\dagger}\cdot v_{2}\left[  1\right]
\right\Vert =\left\Vert v_{3}\left[  3\right]  ^{\dagger}\cdot v_{2}\left[
1\right]  \right\Vert =\frac{1}{\sqrt{3}}%
\end{equation}
only for the given value of $s$ as in the previous case.
For the pair $el_{2}\left(  16\right)  $ and $el_{3}\left(  19\right)  $ the
mixing matrices for $\varphi=0$ are given by%
\[
U_{1}=%
\begin{bmatrix}
0.80217e^{-1.82071i} & 0.57735e^{-0.868576i} & 0.152283e^{-3.1252i}\\
0.36647e^{2.60604i} & 0.57735e^{0.0831114i} & 0.729634e^{-0.0791203i}\\
0.471405e^{-3.09347i} & 0.57735e^{1.25755i} & 0.666667e^{-2.24541i}%
\end{bmatrix}
\]%
\[
U_{2}=%
\begin{bmatrix}
0.80217e^{-1.82071i} & 0.57735e^{-0.868576i} & 0.152283e^{-3.1252i}\\
0.471405e^{-3.09347i} & 0.57735e^{1.25755i} & 0.666667e^{-2.24541i}\\
0.36647e^{2.60604i} & 0.57735e^{0.0831114i} & 0.729634e^{-0.0791203i}%
\end{bmatrix}
~\cdot
\]
Again, making use of the degenerate subspace  we can  find the range of $\varphi$ values in accordance with the experimental
findings. For  the two examples above  the range of $\varphi$ compatible with the experiment  is  between $[-0.0196, 0.123]$
and is plotted in figure~\ref{fig:2}.

\subsubsection{ The pair $el_{2}\left( 10\right) $, $el_{7}\left( 23\right) $ }
We proceed now to an example which involves seventh-order elements of $PSL_2(7)$.
We take the pair $el_{2}\left( 10\right) $, $el_{7}\left( 23\right) $ which is represented by the matrices
\[
li_{2}\left( 10\right) =%
\begin{bmatrix}
r_{1} & -r_{2} & -r_{3} \\
-r_{2} & r_{3} & r_{1} \\
-r_{3} & r_{1} & r_{2}%
\end{bmatrix},
\ li_{7}\left( 23\right) =%
\begin{bmatrix}
r_{1}e^{\frac{4\pi i}{7}} & r_{2}e^{\frac{\pi i}{7}} & r_{3}e^{-\frac{5\pi i%
}{7}} \\
r_{2}e^{-\frac{3\pi i}{7}} & r_{3}e^{-\frac{6\pi i}{7}} & r_{1}e^{\frac{2\pi
i}{7}} \\
r_{3}e^{-\frac{3\pi i}{7}} & r_{1}e^{-\frac{6\pi i}{7}} & r_{2}e^{\frac{2\pi
i}{7}}%
\end{bmatrix}%
~\cdot
\]%
The eigenvectors of  $li_{2}\left( 10\right) $ are%
\[
v_{2}\left[ 1\right] =%
\begin{bmatrix}
\sqrt{\frac{2}{3}-s^{2}} \\
\frac{\sqrt{3}}{2}s+\frac{1}{2}\sqrt{\frac{2}{3}-s^{2}} \\
-\frac{\sqrt{3}}{2}s+\frac{1}{2}\sqrt{\frac{2}{3}-s^{2}}%
\end{bmatrix},\,
\ v_{2}\left[ 2\right] =\frac{1}{\sqrt{3}}%
\begin{bmatrix}
-1 \\
1 \\
1%
\end{bmatrix},\,
\ v_{2}\left[ 3\right] =%
\begin{bmatrix}
s \\
\frac{1}{2}s-\frac{\sqrt{3}}{2}\sqrt{\frac{2}{3}-s^{2}} \\
\frac{1}{2}s+\frac{\sqrt{3}}{2}\sqrt{\frac{2}{3}-s^{2}}%
\end{bmatrix}%
\]
corresponding to the eigenvalues $1$,$-1$, $-1$ repectively. Here $s$ is
given by
\be
s=\sqrt{\frac{2}{3}}\frac{\left( 2+3 r_{3}\right) }{\sqrt{3\left(1+
r_{1}-r_2\right) ^{2}+\left(2+3 r_{3}\right) ^{2}}}\approx 0.732
\cdot
\label{sval7}
\ee
The normalized eigenvectors of $li_{7}\left(
23\right) $ are given by
\[
v_{7}\left[ 1\right] =%
\begin{bmatrix}
r_{3}e^{\frac{4\pi i}{7}} \\
r_{1}e^{-\frac{5\pi i}{7}} \\
r_{2}e^{\frac{5\pi i}{7}}%
\end{bmatrix},\,
v_{7}\left[ 2\right] =%
\begin{bmatrix}
r_{2}e^{\frac{6\pi i}{7}} \\
r_{3}e^{-\frac{3\pi i}{7}} \\
-r_{1}%
\end{bmatrix},\,
v_{7}\left[ 3\right] =%
\begin{bmatrix}
-r_{1} \\
r_{2}e^{-\frac{2\pi i}{7}} \\
r_{3}e^{-\frac{6\pi i}{7}}%
\end{bmatrix}%
\]%
and correspond to the eigenvalues  $e^{\frac{6\pi i}{7}}$, $e^{\frac{10\pi i%
}{7}}$, $e^{\frac{12\pi i}{7}}$respectively. It turns out that the diagonalizing
matrices of all order seven  elements  for some unclear reason can
be written as latin square matrices which, however, do not constitute
elements of the group.  The mixing matrices  compatible with data are
\begin{equation}
U_1=%
\begin{bmatrix}
v_{7}\left[ 1\right] ^{\dagger }, & v_{7}[2]^{\dagger }, & v_{7}\left[ 3\right]
^{\dagger }%
\end{bmatrix}\cdot
\begin{bmatrix}
v_{2}\left[ 1\right],  & \widetilde{v}_{2}\left[ 3\right],  & \widetilde{v}_{2}%
\left[ 2\right]
\end{bmatrix}
\end{equation}
\begin{equation}
U_2=%
\begin{bmatrix}
v_{7}\left[ 1\right] ^{\dagger }, & v_{7}[3]^{\dagger }, & v_{7}\left[ 2\right]
^{\dagger }%
\end{bmatrix}\cdot
\begin{bmatrix}
v_{2}\left[ 1\right],  & \widetilde{v}_{2}\left[ 3\right],  & \widetilde{v}_{2}%
\left[ 2\right]
\end{bmatrix}
~\cdot
\end{equation}%
Note that for $\varphi =0$ the mixing matrices are completely out. However, for $
\phi =\frac{2\pi }{7}$  we get
\[
U_{1}=%
\begin{bmatrix}
0.814857e^{-\frac{3\pi i}{7}} & 0.558406e^{\frac{\pi i}{14}} & 0.15532e^{\frac
	{\pi i}{14}}\\
0.362646e^{\frac{4\pi i}{7}} & 0.700416e^{\frac{\pi i}{14}} &
0.614741e^{-\frac{13\pi i}{14}}\\
0.452212e^{-\frac{3\pi i}{7}} & 0.444523e^{-\frac{13\pi i}{14}} &
0.773242e^{--\frac{13\pi i}{14}}%
\end{bmatrix}
\]%
\begin{equation}
U_{2}=%
\begin{bmatrix}
0.814857e^{-\frac{3\pi i}{7}} & 0.558406e^{\frac{\pi i}{14}} & 0.15532e^{\frac
	{\pi i}{14}}\\
0.452212e^{-\frac{3\pi i}{7}} & 0.444523e^{-\frac{13\pi i}{14}} &
0.773242e^{--\frac{13\pi i}{14}}\\
0.362646e^{\frac{4\pi i}{7}} & 0.700416e^{\frac{\pi i}{14}} &
0.614741e^{-\frac{13\pi i}{14}}%
\end{bmatrix}
~.
\end{equation}
The range of $\delta \varphi \equiv \varphi -\frac{2\pi}7 $ that falls within the experimental bounds is very narrow, and is given by
\be
 -0.0028 < \delta \varphi< +0.03\label{phibound}
\ee
\begin{figure}[t]
	\begin{center}
		\includegraphics[width=0.465\textwidth,angle=0,scale=0.9]{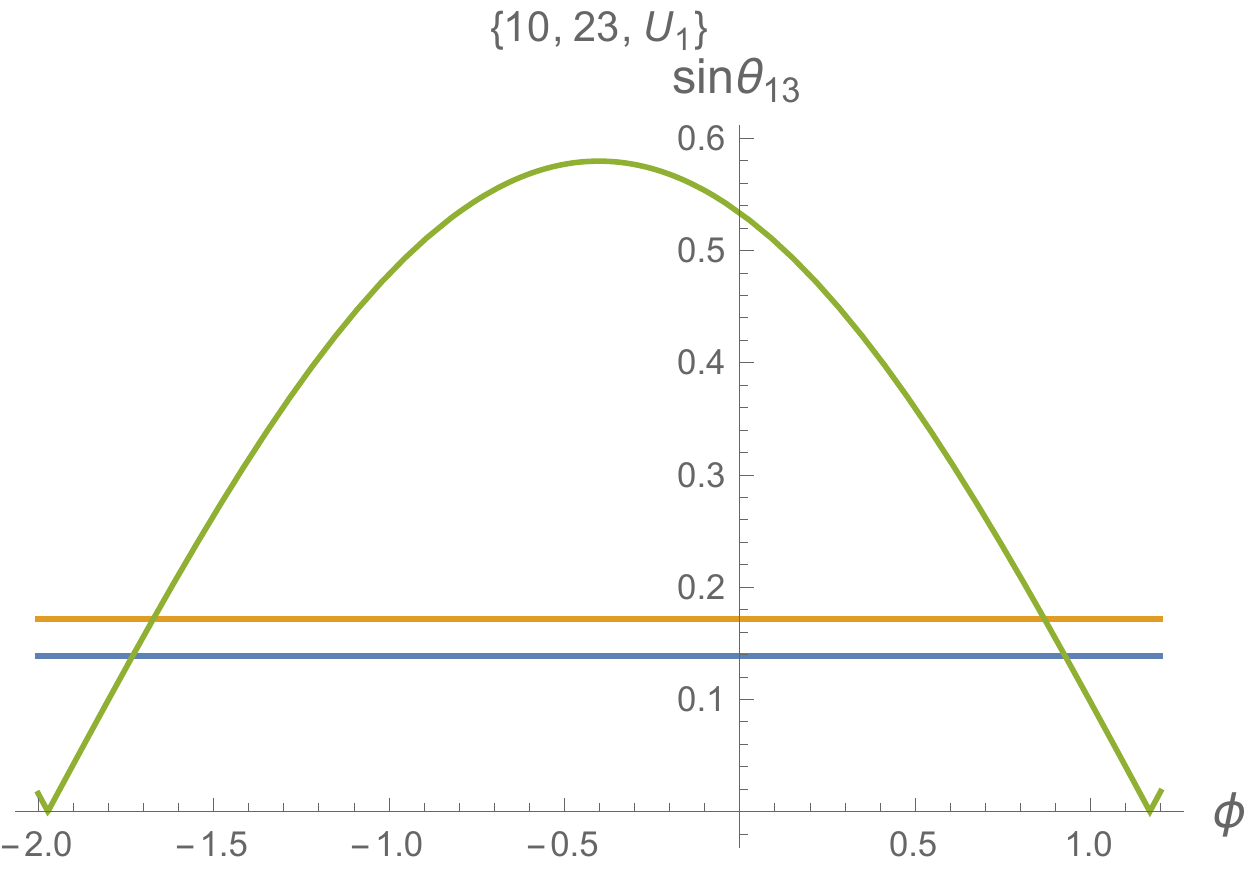}
		\hspace{0.1cm}
		\includegraphics[width=0.465\textwidth,angle=0,scale=0.9]{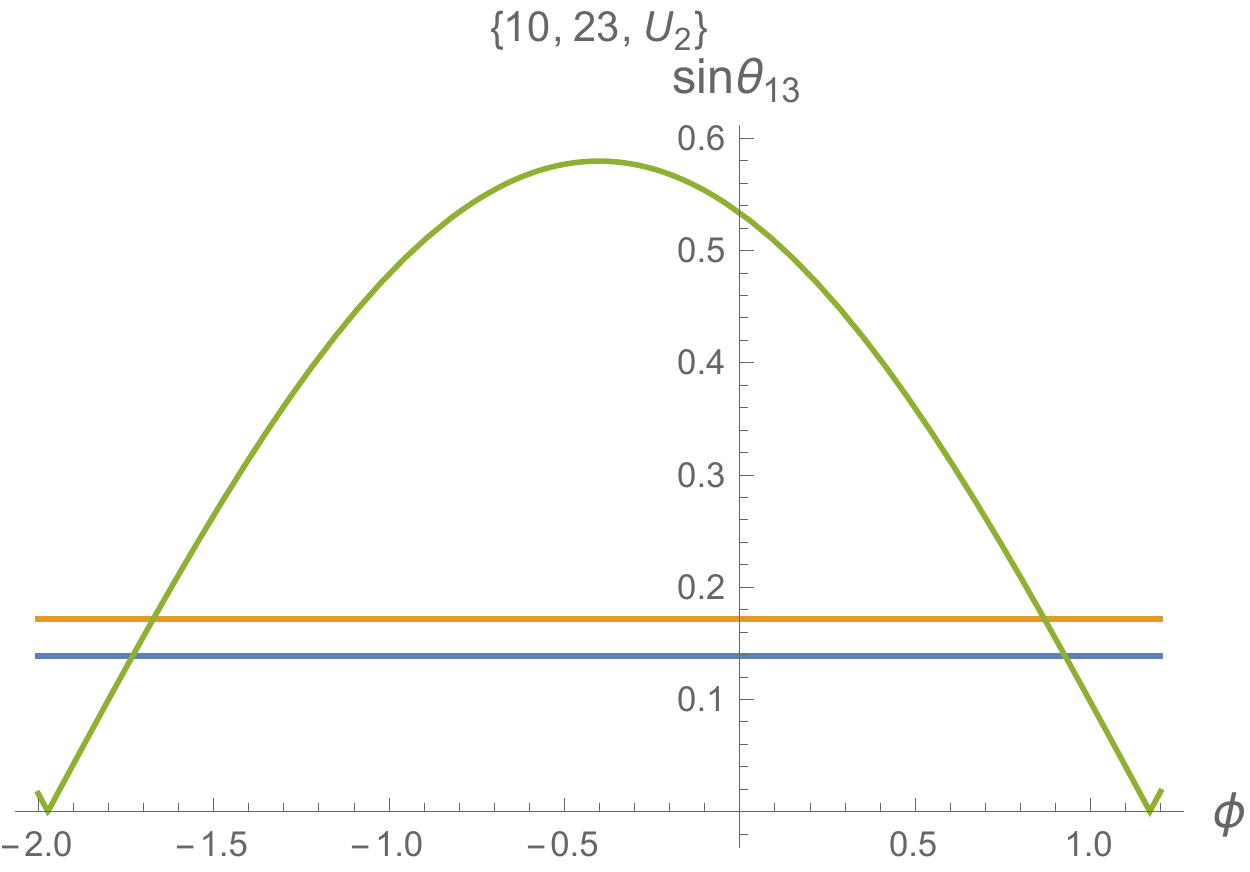}
	\end{center}
	\caption{
	Plots show  the experimentally compatible range of $\sin\theta_{13}$ as function of $\varphi$, for example 3 involving $PSL_2(7)$ elements of order 7.
Orange and blue lines define the experimental bounds.	}
	\label{fig:3}
\end{figure}

\section{Summary and Conclusions}

The last couple of decades, a substantial amount of research in physics beyond the Standard Model  has been devoted to
interpret the lepton mixing matrix, and in particular, the neutrino data. A rather established approach to this  task is
to postulate invariance of the Yukawa lagrangian under some suitable finite group.  Remarkably, such symmetries  appear
naturally in a wide class of extensions of the Standard Model emerging  in the framework of  String and F-theory constructions.
Given these facts and the continuing interest on these issues as well as the considerably wide applications of the discrete
groups in phenomenological models,  in this article, we focused our investigations on the projective linear group $PSL_2(7)$.
This group is a simple discrete subgroup of $SU(3)$, and the largest one possessing  three-dimensional unitary representations.
Therefore it is a suitable candidate for a discrete flavor symmetry of the effective  theory. However, despite  its  interesting features,
 its implications in low energy phenomenology  have not been widely explored, partially because of the apparent complicated
  structure of the representations of its elements.

In this work we generate viable textures of charged-leptons neutrino mixing matrices by assuming that both types of matrices
commute with some element of the $PSL_2(7)$ group. A tedious calculation reveals that the basic hypothesis is correct and it
is valid for a number of group elements which are given in the context. It turns out that the neutrino mass matrix can only
commute with order $2$ group elements while the charged leptons mass matrix can commute with both order $3$ and order $7$ elements.
The results indicate there are only two types of matrices for the $\left(  el_{3},el_{2}\right)$ combinations and also only two
for the $\left(  el_{7},el_{2}\right)$ combination. The eigenvector degeneracy of the $el_{2}$ elements (parametrized by a single
parameter, the angle $\varphi$)   may change somehow  the values of the mixing angles. It appears though that the allowed values of
the free parameter $\varphi$ are very strongly centered  around a fixed value which is $\varphi=0$ for the  $\left(  el_{3},el_{2}\right) $
combination and $\varphi=\frac{2\pi}{7}$ for the  $\left(  el_{7},el_{2}\right)$. The value  $\varphi=0$  suggests that nature prefers
the eigenvector $\frac{1}{\sqrt{3}}\left( -1,\, 1,\,  1\right) $, which is independent of the values $\rho_{1},\rho_{2},\rho_{3}$ which
characterize the group $PSL_{2}\left(  7\right)  $. In the more general case, this eigenvector is replaced by  $\frac{1}{\sqrt{3}}{\left(e^{ic_{3}},  e^{i\left(  c_{3}-c_{2}\right)  }, 1\right)} $ which is again independent of the values $\rho_{1},\rho_{2},\rho_{3}$ and the results obtained are similar to those presented here. Note that the algebra of the  pairs $\left(  el_{3},el_{2}\right)  $ generate various
$A_{4}$ subgroups of ~$PSL_{2}\left(  7\right)  $. Probably this the reason for the appearance of the resulting mixing matrix as a
generalization of the tri-bi-maximal mixing~\cite{Harrison:2002er}, i.e. the moduli of the middle column elements are equal to $\frac{1}{\sqrt{3}}$
a phenomenon which is known to be connected with an $A_{4}$ symmetry. Here, this simplification occurs non trivially because the form of the elements is very
complicated. As for the $\left(  el_{7},el_{2}\right)  $ pairs, no $PSL_{2}\left(  7\right)  $ subgroup is generated a fact that leaves more
space to  the middle column elements to arrange themselves. Note that the allowed values of $\varphi$ are centered around the central value $\varphi
=\frac{2\pi}{7}$ which is  the phase associated with the basic seventh root of unity $e^{i\frac{2\pi}7}$.


\vspace{2cm}
{\bf Acknowledgement}. NDV would like to thank Theory Division at CERN, for kind hospitality during the final stages of this work.

\appendix

\newpage

\section{Appendix}

The following tables depict the correspondence between the $PSL_{2}\left(
7\right)  $ elements as calculated and enumerated by $GAP$ and their
corresponding distribution among conjugacy classes used in the text.

\begin{itemize}
\item Character $-1$%

\[%
\begin{tabular}
[c]{lllllllll}\hline
$el_{2}$ & $1$ & $2$ & $3$ & $4$ & $5$ & $6$ & $7$\\
$GAP$ & $119$ & $120$ & $121$ & $122$ & $123$ & $124$ & $125$\\\hline
$el_{2}$ & $8$ & $9$ & $10$ & $11$ & $12$ & $13$ & $14$\\
$GAP$ & $140$ & $141$ & $142$ & $143$ & $144$ & $145$ & $146$\\\hline
$el_{2}$ & $15$ & $16$ & $17$ & $18$ & $19$ & $20$ & $21$\\
$GAP$ & $161$ & $162$ & $163$ & $164$ & $165$ & $166$ & $167$\\\hline
\end{tabular}
\ \ \ \ \ \ \
\]

\item Character $0$
\[
\begin{tabular}{lllllllll}
\hline
$el_{3}$ & $1$ & $2$ & $3$ & $4$ & $5$ & $6$ & $7$ & $8$ \\
$GAP$ & $1$ & $2$ & $3$ & $4$ & $5$ & $6$ & $7$ & $8$ \\ \hline
$el_{3}$ & $9$ & $10$ & $11$ & $12$ & $13$ & $14$ & $15$ & $16$ \\
$GAP$ & $9$ & $10$ & $11$ & $12$ & $13$ & $14$ & $51$ & $52$ \\ \hline
$el_{3}$ & $17$ & $18$ & $19$ & $20$ & $21$ & $22$ & $23$ & $24$ \\
$GAP$ & $53$ & $54$ & $55$ & $56$ & $57$ & $58$ & $59$ & $60$ \\ \hline
$el_{3}$ & $25$ & $26$ & $27$ & $28$ & $29$ & $30$ & $31$ & $32$ \\
$GAP$ & $61$ & $62$ & $63$ & $64$ & $72$ & $73$ & $74$ & $75$ \\ \hline
$el_{3}$ & $33$ & $34$ & $35$ & $36$ & $37$ & $38$ & $39$ & $40$ \\
$GAP$ & $76$ & $77$ & $78$ & $97$ & $98$ & $99$ & $100$ & $101$ \\ \hline
$el_{3}$ & $41$ & $42$ & $43$ & $44$ & $45$ & $46$ & $47$ & $48$ \\
$GAP$ & $102$ & $103$ & $126$ & $127$ & $128$ & $129$ & $130$ & $131$ \\
\hline
$el_{3}$ & $49$ & $50$ & $51$ & $52$ & $53$ & $54$ & $55$ & $56$ \\
$GAP$ & $132$ & $133$ & $134$ & $135$ & $136$ & $137$ & $138$ & $139$ \\
\hline
\end{tabular}%
\]

\item Character $+1$%
\[%
\begin{tabular}
[c]{llllllll}\hline
$el_{4}$ & $1$ & $2$ & $3$ & $4$ & $5$ & $6$ & $7$\\
$GAP$ & $16$ & $17$ & $18$ & $19$ & $20$ & $21$ & $22$\\\hline
$el_{4}$ & $8$ & $9$ & $10$ & $11$ & $12$ & $13$ & $14$\\
$GAP$ & $23$ & $24$ & $25$ & $26$ & $27$ & $28$ & $29$\\\hline
$el_{4}$ & $15$ & $16$ & $17$ & $18$ & $19$ & $20$ & $21$\\
$GAP$ & $30$ & $31$ & $32$ & $33$ & $34$ & $35$ & $36$\\\hline
$el_{4}$ & $22$ & $23$ & $24$ & $25$ & $26$ & $27$ & $28$\\
$GAP$ & $37$ & $38$ & $39$ & $40$ & $41$ & $42$ & $43$\\\hline
$el_{4}$ & $29$ & $30$ & $31$ & $32$ & $33$ & $34$ & $35$\\
$GAP$ & $44$ & $45$ & $46$ & $47$ & $48$ & $49$ & $50$\\\hline
$el_{4}$ & $36$ & $37$ & $38$ & $39$ & $40$ & $41$ & $42$\\
$GAP$ & $79$ & $80$ & $81$ & $82$ & $83$ & $84$ & $85$\\\hline
\end{tabular}
\ \
\]

\item Character $-\frac{1}{2}-i\frac{\sqrt{7}}{2}$%
\[%
\begin{tabular}
[c]{lllllllll}\hline
$el_{7}$ & $1$ & $2$ & $3$ & $4$ & $5$ & $6$ & $7$ & $8$\\
$GAP$ & $65$ & $66$ & $67$ & $68$ & $69$ & $70$ & $71$ & $86$\\\hline
$el_{7}$ & $9$ & $10$ & $11$ & $12$ & $13$ & $14$ & $15$ & $16$\\
$GAP$ & $87$ & $88$ & $89$ & $90$ & $91$ & $92$ & $93$ & $94$\\\hline
$el_{7}$ & $17$ & $18$ & $19$ & $20$ & $21$ & $22$ & $23$ & $24$\\
$GAP$ & $95$ & $96$ & $104$ & $105$ & $106$ & $107$ & $108$ & $109$\\\hline
$el_{7}$ & $25$ & $26$ & $27$ & $28$ & $29$ & $30$ & $31$ & $32$\\
$GAP$ & $110$ & $111$ & $112$ & $113$ & $114$ & $115$ & $116$ & $117$\\\hline
$el_{7}$ & $33$ & $34$ & $35$ & $36$ & $37$ & $38$ & $39$ & $40$\\
$GAP$ & $118$ & $147$ & $148$ & $149$ & $150$ & $151$ & $152$ & $153$\\\hline
$el_{7}$ & $41$ & $42$ & $43$ & $44$ & $45$ & $46$ & $47$ & $48$\\
$GAP$ & $154$ & $155$ & $156$ & $157$ & $158$ & $159$ & $160$ & $168$\\\hline
\end{tabular}
\ \ \ \
\]

\end{itemize}

\section{\bigskip The general eigenvectors of the matrix $M$.}

Given $\rho$ the three distinct eigenvalues of the matrix $M$ , the
non-normalized  eigenvectors are given by the expression%

\begin{equation}
v=\left(
e^{ic_{3}}r_{2}\left(  1+\rho~e^{i\left(  c_{1}+c_{5}\right)  }\right),\,
e^{i\left(  c_{3}-c_{2}\right)  }\left[  \rho^{2}e^{i\left(  c_{1}
	+c_{5}\right)  }-\rho\left(  r_{2}+r_{1}e^{i\left(  2c_{1}+c_{5}\right)
}\right)  \right]  +r_{3}e^{ic_{1}},\, r_{1}\left(  \rho+e^{ic_{1}}\right)
\right)
\cdot\nn
\end{equation}
When normalized, while the $el_{2}$ and $el_{3}$ elements do not produce
anything worth mentioning the $el_{7}$ eigenvectors produce diagonalizing
matrices which are latin squares. All the phases can be exactly calculated
however, the trace of the resulting matrix to the best of our knowledge does
not correspond to a known group character so this tantalizing result must
remain a curiosity for the time being.

\end{document}